# A Testable Theory for The Emergence of the Classical World

Stuart Kauffman[1] and Sudip Patra[2]

[1] Institute for Systems Biology
[2] Jindal Global University, CEASP

**Abstract**

The transition from the quantum to the classical world is not yet understood. Here we take a new approach. Central to this is the understanding that measurement and actualization cannot occur except in some specific basis. But we have no established theory for the emergence of a specific basis. Our framework entails the following: i. Sets of N entangled quantum variables can mutually actualize one another. ii. Such actualization must occur in only one of the $2^N$ possible bases. iii. Mutual actualization progressively breaks symmetry among the $2^N$ bases. iv. An emerging "amplitude" for any basis can be amplified by further measurements in that basis, and it can decay between measurements. v. The emergence of any basis is driven by mutual measurements among the N variables, and decoherence with the environment. Quantum Zeno interactions among the N variables mediates the mutual measurements. vi. As the number of variables, N, increases, the number of Quantum Zeno mediated measurements among the N variables increases. We note that decoherence alone does not yield a specific basis. vii. Quantum ordered, quantum critical, and quantum chaotic peptides that decohere at nanosecond versus femtosecond time scales, can be used as test objects. viii. By varying the number of amino acids, N, and the use of quantum ordered, critical, or chaotic peptides, the ratio of decoherence to Quantum Zeno effects can be tuned. This enables new means to probe the emergence of one among a set of initially entangled bases via weak measurements after preparing the system in a mixed basis condition. ix. Use of the three stable isotopes of carbon, oxygen, and nitrogen, and the five stable isotopes of sulfur, allows any ten atoms in the test protein to be discriminably labeled and basis emergence for those labeled atoms can be detected by weak measurements. We present an initial mathematical framework for this theory, and we propose experiments.

**Keywords:** quantum to classical transition, collectively actualizing sets, basis symmetry breaking, quantum ordered, critical, chaotic peptides, Quantum Zeno effects,

decoherence, tunable ratio of decoherence to Quantum Zeno effect, entangled bases, weak measurements, isotope labeling.

**Introduction**

"Is the moon there when we are not looking?" was Einstein's quip. We wish to approach the emergence of a classical world from the barest foundations of quantum theory: N entangled variables, with not even a basis chosen for measurement. Here we take measurement to be real and to constitute an 'actualization' of the quantum state to yield Boolean true false variables. We have at our disposal in the interactions among these N variables: decoherence, recoherence, the Quantum Zeno Effect, and actualization. The "Dud Bomb" work, [1], supports our new proposal that actualization interactions among the N variables can occur. We propose further that this enables the variables to <u>collectively "look at" and mutually actualize one another</u>. We will call such a set of coupled variables a Collectively Actualizing Set, CAS. In such a CAS, the frequency with which each variable will be measured increases with the number of variables, most simply in a linear way.

There is a parallel to a similar idea about the origin of life via 'Collectively Autocatalytic Sets' seen as molecular fossils in ancient prokaryotes, [2] Such small molecule CAS instantiates Rovelli et al.'s 'Stable Facts', [3], as the molecules explode into collective existence.

Actualization itself can only occur in some specific basis among the $2^N$ bases. In the present article, we propose that a set of N entangled variables with no chosen measurement basis among the $2^N$ possible bases, can break this symmetry by co-actualization in some one specific basis. Repeated actualizations in the same single basis amplifies the preference for that specific basis among the N entangled variables. As a preference to actualize in this one basis emerges, and is shared among the N entangled variables, the variables progressively become ever more stable facts with respect to one another.

The emergence of these 'ever more stable facts' drives the symmetry breaking toward one among the $2^N$ bases is chosen. Following the relational quantum mechanics framework (RQM) suggested first by Rovelli, and developed over decades [3,4,5,6], we are to consider unstable relative facts of one variable which need to be labelled against at least one other variable in the system from whose perspective such facts are actualized. But there are also stable facts, each of which does not require such labelling, and appear to some other variables in the systems (or 'observers') as an invariant fact.

Stable facts are brought into being by environmental decoherence. In our expanded framework we consider an ongoing interplay of decoherence and the Quantum Zeno

Effect. We propose that the classical world emerges in Collectively Actualizing Sets whose variables undergo both environmental decoherence and mutually induce a QZE effect with one another. These interactions lead to a broken symmetry such that a specific basis emerges, and the system is repeatedly actualized in the same state in that basis by very rapid Quantum Zeno effects among the N variables. This is to be 'the classical world'.

The dynamic is complex: i. Repeated actualization drives the Quantum Zeno Effect [7, 8], in which the wave function transiently freezes to a single state in some specific basis then rebounds quadratically in time. ii. Decoherence [9,10] by itself is not measurement. It yields 'classical FAPP' with classical probabilities down the main diagonal and no way to pick among them. iii. The Quantum Zeno Effect must occur in some specific basis. iv. Decoherence tends to suppress all bases other than position and momentum. v. The Quantum Zeno effect and decoherence mutually suppress one another [11].

Given mutual suppression between decoherence that leads only to classical probabilities and the QZE that freezes the wave function to a single state in one specific basis, we here explore for the first time the joint dynamics of decoherence and QZE in a set of N entangled variables that form a Collectively Actualizing Set.

We study our system in the context of a spin glass Hamiltonian that can model a peptide folding energy landscape, [12] Does this joint dynamics evolve to a specific basis, and, within it, hover at a specific state a by repeated CAS measurements? Is such a state 'the classical world'? How is it affected by the number of variables, N, in the Collectively Actualizing Set? Increasing N should increase the frequency of QZE, hence tune the ratio of decoherence to QZE. How is this classical outcome affected by the ratio of decoherence to QZE?

Remarkably, our predictions can now be tested using novel, constructable short ring or linear peptides with a defined tunable number, N, of amino acids that are quantum 'ordered', 'critical' or 'chaotic' [13,14]. Ordered and critical peptides decohere power law slowly over a nanosecond. Chaotic peptides decohere exponentially fast over a femtosecond. Thus, we can study the relation between ratio of decoherence, and a recurrent Quantum Zeno Effect whose frequency is positively correlated to the number of amino acids, on the emergence of one among perhaps several bases.

Naively we expect larger peptides to become 'classical' faster than shorter peptides. We here explore how this may depend upon the 'ratio' of decoherence to QZE as affected by short or long peptides that decohere exponentially fast versus power law slow. This can be tested by using weak measurements. We can assess the time course of an emerging 'amplitude' for a given basis in a system prepared in a superposition of two or more bases. Interactions between two or more emerging bases can also be studied using systems prepared in an initial superposition for those two or more bases, as further explained below.

Using the three stable isotopes of carbon, oxygen, and nitrogen, and the five stable isotopes of sulfur, any ten detectably isotope labeled atoms can be located at any location in such quantum ordered, critical, or chaotic peptides, allowing precise analysis of the emergence of bases and the classical world.

These ideas and their initial mathematical formulation and experimental approaches outlined here, constitutes our "testable theory for the emergence of the classical world".

> The paper is organized thus: Section 1 elaborates the conceptual framework, Section 2 provides the experimental set up for testing our central questions, and Section 3 concludes with further discussion.

**Section 1**

We suggest here verifiable experiments, establishing the salient features as follows: We start with classifying the environment as a heterogeneous ensemble of mutually actualizing sets composed of dissipative quantum systems. Hence one such mutually actualizing set can be perceived as a system to be 'measured' by the rest of the ensemble. Here, by measurement we mean interaction, rather than any special metaphysical position provided to an 'observer'. We retain the essence of relational QM by suggesting any variables within a mutually actualizing set can work as an observer in relation to others. Hence all states generated in such interactions are relative states, and such interactions are 'relative facts'. Environmental decoherence can lead to 'stable facts'; such facts need not be labelled against any particular system. Generally, as the extensive literature [16] suggests, the environment 'measures' a quantum dissipative system, where two scenarios may emerge. One is standard decoherence where via Einselection only stable states (or stable facts here) survive and emerge as 'classical' observables in two preferred basis: The pointer bases of position and momentum. Such a process which is generated by entanglement with the environmental degrees of freedom of the system, can be approximated as anti- QZE.
The other possible case, there can be continuous projective measurement on the system by the environment, and in the limit, survival probability of the system's initial state (say a pure state) tends to unity, which is QZE. It is the intermediate case, which is of interest to us here.

Our framework differs in a fundamental and new way: We suggest that the mutually actualizing sets themselves generate many body QZE that are each an actualization in some single basis, while the CAS simultaneously decoheres due to coupling with environment, where such couplings can be non-uniform across such sets/ systems.

The important point is: In the target collectively/ mutually actualizing set of the subsystems no preferred basis is yet chosen while many body QZE happens, where, as the first approximation, the rest of the environment has a definite basis. Hence decoherence driven by environmental coupling might generate a pointer bases for the target system.

The total Hamiltonian is expressed as $H = H(S) + H(E) + H(Int)$. The symbols carry usual meanings. H is the total system Hamiltonian, H(S) is the system Hamiltonian, H(E) is the Hamiltonian representing the environment, and H(int) is the interaction Hamiltonian. *Here S denotes system, E denotes environment and Int denotes coupling.* For states generated via the mutual actualization (many body QZE) to emerge as stable facts/ pointer states, such states need to be Eigen states of the H in general. Here however, we have three regimes, first where H(S) is the dominating part, second, where H (int) is the dominating part, and then the third intermediate regime.
We may simplify the scenario by assuming that to start with the system's state and the environment state is in a tensor product state, on which the evolution happens. Hence, we may calculate the reduced density matrix of the system by tracing over the combined state for environmental degrees of freedom. Such a reduced density matrix then follows the von Neumann equation [17].

Many Body QZE

Interest in many body QZE is rather a recent phenomenon [18], where many body QZE generating entanglement phase shifts has been studied. Our collectively/ mutually actualizing set of systems based many- body QZE differs in some points from the extant framework. We lay down the salient features of the mechanism below:

- Continuous (in the limit) projective/ POVM measurements simultaneously acting on the system of many bodies (for example, all lattice points), in effect localizes the evolution of the composite state. However, in our case the frequency with which a given subsystem is measured is proportional to the number of subsystems in the collectively actualizing set, so as that number goes up any subsystem is measured proportionally more frequently.
- In the extant literature we have continuous (in limit) projective or POVM on the entangled many-body dynamics. Such a measurement process would generate 'entanglement' phase shifts, which might be expressed through suitable

- Hamiltonians of the system, for example that of standard Ising model or a spin glass model [19].
- In our framework, patterns of entanglement among the same set of variables within a collectively actualizing set change, due to successive actualizations.
- POVM measures rather than projective measures are weak, but such measures can also create stochastic 'back action' on the mutually act set, also generating entanglement phase shifts.
- Such phase shifts are observed as single quantum Many body states.
- Hence, we preserve the overall competition between unitary many body dynamics and the stochastic measurements of single members of an entangled subset of the CAS.
- Authors have observed [20,21] that such processes involve degrees of freedom which do not commute with each other, for example different components of spin (directions). Starting with a simple one dimensional Ising model, such many body QZE can produce a frozen / localized many body state when the coupling is above critical threshold strength. As that threshold is approached, a behaviour resembling quantum criticality emerges.
- Explicitly, a quantum many body dynamics $|\varphi(t+dt)> = MU|\varphi(T)>$, where U is the dynamics generator / unitary operator $U = \exp(-i\,Ht)$, *and where H is the system H*. H can be either Ising H or Spin glass H. For example, in the Ising Model:
- $H = J\sum_{i=1}^{L-1} \sigma_i\,\sigma_{i+1}$ Where sigmas are the Pauli matrices working on the ith lattice site, and the sigmas go over x, y, z.
- $M$ is the measurement operator, representing the simultaneous measurement carried over all sites, with a probability, p, per site. Such a measurement can be represented by the KRAUSS operator.
- $M = $ *TENSOR product over $M_i$, where Mi s are composed of the one D projectors on the sites. Using the K operator* does not carry out sequential measurements since it represents simultaneous measurement on all sites of the Lattice. We might begin using such a picture as an approximation. Later, we need sequential measurements of variables in any one of the entangled subsets of variables in the CAS.
  *The probability of M measures on site i is proportional to n.* Again, captures our assumption that frequency with which a subsystem gets actualized is proportional to the number of surrounding subsystems.

Tradeoff between QZE and de-coherence: interaction Hamiltonian plays central role in both De-coherence framework as well as stochastic quantum hydrodynamics framework, which we describe later. However, since we are interested how mutual actualization (approximated by many body QZE) helps in emergence of specific basis, while the de-coherence is present in the background environment, an intermediate regime where neither QZE nor de-coherence dominates each other is of importance to the framework. Since by definition QZE (as we describe later which can be thought as a sequence of weak measurements in general) would freeze unitary evolution, whereas de-coherence is an approximate process of entanglement of the system (here for example the collectively actualizing set of bodies) with the environmental degrees of freedom, to generate 'improper mixed states'.

**Section 2**

2.1 A Prospective Maximum Amplitude for Each Basis as a Function of the Rate of QZE

This is new physics. We do not know if, and if so, how amplitudes for the different possible bases may emerge, how repeated actualizations in any single bases may affect the emergence of that basis, how these different emerging bases may interact, or if, between actualization events, an emerging basis decays, if so at what rate, and how decoherence by coupling to an environment alters all this.

Our fundamental proposal envisions a Collectively Actualizing Set, CAS, in which the N variables interact – actualize episodically with one another. Each actualization is in some single basis. We propose that when one among an entangled subset of the N actualizes in some basis, the 'commitment to' that basis, or 'amplitude for' that basis, increases in a stepwise fashion by a fixed increase in amplitude for all the entangled variables in that subset. Importantly, during temporal intervals in which no actualization events occur among the subset of the N entangled variables, we propose that the commitment to, or amplitude for that basis decays at some fixed exponential rate for the entire subset of the N entangled variables.

These assumptions imply a relation between the frequency of measurement and the maximum amplitude that can be attained for any basis. Because we have assumed an exponential rate of decay of an amplitude for any basis, and a fixed increase in that amplitude for each actualization event in that basis, an equilibrium maximum amplitude for each rate of measurement, F, must exist where the exponential rate of loss of amplitude for that basis equals the mean rate of increase of amplitude for that bases by actualization events in that basis.

Because we assume that the frequency of episodic interaction – actualization events increase linearly as the number of variables, N, increase, the maximum amplitude attainable for any basis should increase linearly with the number of atoms in the system, hence the number of amino acids in the test peptides.

If we posit that temperature, classical noise, erases any accrued amplitude for any basis, increasing temperature, other variables fixed, should inhibit the emergence of any single basis and the classical world. Given the collective dynamics of our system, the effect of increasing temperature may appear as a phase transition.

Our experiments below should be able to prove or disprove these qualitative predictions.

We hope to study the emergence of a bias for one among a large set of bases as a function of the ratio of decoherence to the Quantum Zeno effect. We propose to measure the ratio of decoherence to QZE, as it is approximated by the use of a relative entropy measurement, for example relative von Neumann measure. Here we start with a mixed state of one such CAS, when certain individual particles are actualized or they get dis-entangled, the state of the CAS changes, and hence the entanglement entropy measure:

> The von *Neumann* entropy measure can be used to measure the change in the degree of entanglement brought about by the QZE. The ratio of QZE/DECOHERENCE would then be correlated with the oscillation of the von Neumann entropy measure.
> *von Neumann relative entropy*, $NE(\rho||\sigma) = \frac{tr\{\rho ln\rho - \rho ln\sigma\}}{tr\rho}$, where $\rho$ and $\sigma$, are states before and after collective actualization phases.
> Hence NE is greater than or equal to 0, equality holding when $\rho = \sigma$, *unitary invariance*
> $NE(U\rho U^D||U\sigma U^D) = NE(\rho||\sigma)$, here D symbolizes Hermitian conjugate.
> *The maximum divergence occurs when $NE(pure\ state\ ||maximally\ mixed\ state)$* where the maximally mixed state = 1/d, and where d= dim H. Hence, NE(pure state||1/d)= ln d.

2.2. Stochastic quantum hydrodynamics approach (SQHM)

Chiarelli and Chiarelli [22], while exploring the literature on quantum Hydrodynamics, have observed that quantum hydrodynamic equations can play A critical role in describing emergence of the classical world from the underlying quantum domain. The main problem they approach is, however, the fact that the dependence of the dynamics of systems on mass create disconnections in A smooth passage from quantum superposition to classical domain. The authors also observe that many different interpretations (or even different theories than standard QM, for example Bohmian mechanics or different dynamic collapse models which contain different parameters than QM) of QM has been used to reach at a solution. However, our framework IS not directly related to the mass density problem.

Now, some insights from Madelung's version of quantum hydrodynamics help us in explaining our proposed experiments. In Madelung's framework, we have A wave function expressed as $\varphi = |\varphi|e^{2\pi iS/h}$, which is equivalent to the dynamics of a mass density $|\varphi|^2$, *with momemtum* $p_i = \frac{\partial S}{\partial q_i}$. One important observation on Madelung's framework (where the dynamics converges with that of Schrödinger's equation) is that when Planck's constant is set to 0, i.e. we take a classical limit, we recover classical dynamics. Hence, such a framework can be used to describe the transition from the quantum to the classical world. Another natural outcome of the framework is non-locality due to the presence of the quantum potential, which also means the presence of trajectories and physical reality independent of measurement. Now this feature may not be compatible with the collective actualization process we are suggesting here, since the relational ontological view is implicit in our framework, which would imply relative and stable facts. But for existence of both we either need relative states between systems or de-coherence led emergence.

Hence, some insights we may draw from Madelung's stochastic quantum hydrodynamics (SQHM) are as below:

1. In SQHM [22] the range of the non-local quantum potential depends on the strength of the interaction Hamiltonian, which shows that systems sufficiently weakly interacting can generate classical behavior in the macroscopic classical limit. In our framework also we have the central importance of weak measures (which is approximated by weakly interacting subsystems in a collectively actualizing set), and AN interaction Hamiltonian, but we do not have quantum non-local potentials. Hence the process of emergence varies between our and Maudling's SQHM.
2. SQHM and de-coherence are compatible to each other, while SQHM provides limits when a global macroscopic dissipative quantum system acquires classical behavior, de-coherence provides an approximate process of emergence of 'impure mixed' states which mimics classical reality, even when the underlying global system is quantum. The central assumption in case of de-coherence IS that THE recurrence time is absurdly large. Hence as emphasized in the paper also, de-coherence is not a solution to measurement or collapse problem, unlike dynamic collapse theories with non-linear Schrödinger equations. Since SQHM is a special case of Bohmian mechanics, we have included a technical note on the same in the appendix.

**Section 3**

3.1 Experimental Approaches

In order to detect the emergence of a single bases from a larger set of available bases requires an experimental means to initiate a system with more than a single basis available, of which none has as yet been "chosen". A means to accomplish this consists in making use of the capacity to prepare a system in an initial state with two superimposed bases. This can be done among the following five pairs: position – spin, position – polarization, spin-polarization, momentum -spin, momentum – polarization.

We wish to test experimentally a possible relation between the number of variables, N, in a proposed Collectively Actualizing Set, CAS, and the ratio of decoherence to Quantum Zeno effects as the N variables actualize one another, upon the emergence of one or the other of two initially superposed bases, for example, 'position – spin', or 'spin – polarization' by a symmetry breaking among the these 2 possible bases.

We have at our disposal the experimental creation of linear and ring peptides with a tunable number of amino acids per peptide, and choice of which of the standard 20 amino acids occurs at each site in the peptide polymer. Furthermore, by using the three stable isotopes each of carbon, oxygen, nitrogen, and the five stable isotopes of sulfur, we can uniquely isotope label any ten atoms at any ten chosen position in a peptide.

Hence, by timed weak measurements, as described in detail below, we can assess the onset of a bias towards either of the 2 superposed bases, the time course of that emergence, and the stability of that emerging bias toward one of the two bases by a perturbing weak measurement of the other basis.

Our proposed experiments seem to probe entirely new physics. From the hoped-for data, it may be possible to construct a clean theory of collective symmetry breaking among any pair of bases or the entire set of $2^N$ bases of such a quantum system.

We propose using quantum ordered, critical and chaotic peptides as test objects. It is relatively straight forward to test computationally or experimentally, if a given peptide is ordered, critical or chaotic. The distribution of energy differences between adjacent absorption bands falls off exponentially for ordered peptides. If a given peptide is chaotic the distribution is a broad single peak given by random matrix theory. The distribution for critical peptides lies between these two distributions, [13,14].

In order to study the time course of emergence of one among the two superposed bases, requires a single weak measurement to assess the amplitude for a given basis at different time intervals after time 0. To study the interaction of one emerging basis if the other basis is perturbed by a weak measurement requires only one additional weak measurement. That is, if we have already established the time course of the emergence of either of the two bases alone by single measurements at increasing times after time T 0, we can, in a separate experiment, test the effects of weak measurement of the one basis on the emergence of the other basis. The two bases may emerge independently of one another. Or the emergence of one basis may inhibit or enhance the emergence of the other basis. If we can do two measurements during the available time, we can test the effect of perturbing one emerging basis on the emergence of the other basis in a single experimental system.

Thus, slower time scale of decoherence in ordered and critical may better allow study of the temporal emergence of bases choice and the interactions between the emerging bases.

We stress that our mathematical framework, extended to study non-synchronous projective measurements among the N variables, and weak measurements is needed for more precise predictions.

3.2 More Experimental Details

The basic experimental approach is to create libraries of ordered, critical and chaotic ring or linear peptides of tunable number of amino acids from 2 to perhaps 100 amino acids or more. Ring peptides are more confined in their structure than linear peptides, but both have more flexible degrees of freedom than crystals such as C 60 Buckyballs. The flexibility will increase with the number of amino acids in the ring or linear peptide. As noted, a single amino acid has, on average, 10 atoms. We can hope to study single amino acids, 10 atoms which is less than the C 60 Buckyballs that still show inference, to polypeptides of 100 amino acids and1000 atoms. Presumably "classicality" increases as the number of atoms in the system increases.

Decoherence in peptides and proteins as a function of time can be assessed by known techniques, such as those used in studying light harvesting molecules, [22,23,24]. It is not, at present, clear how to directly study the intensity of Quantum Zeno effects within a peptide, however it is reasonable to propose that if the atoms in a peptide form a collectively actualizing set, that QZE interactions will increase monotonically with the number of atoms in the system. This should be directly testable.

Some measures we can think of are based on the literature of quantum thermodynamics as follows: On one hand we have through QZE on entangled N elements (approximation of collective actualization process) the emergence of certain variables in any one basis among $2^N$ bases. On the other hand, we have the emergence of preferred pointer basis, position and momentum, via decoherence. Because we can tune the ratio of QZE and decoherence by use of ordered, critical and chaotic peptides, we can study the effect of tuning this ratio on the trade-off between an emergence of a pointer basis via decoherence alone and the possible emergence of other bases by the QZE and our proposed symmetry breaking.

In short, we have identified some ratios earlier for the relative intensity of many body QZE to decoherence, hence in the intermediate regime when neither of the two processes significantly dominate each other, there are some dynamics which can be studied experimentally.

Generally, if two states $|\varphi_1>$ and $|\varphi_2>$ are perfectly distinguishable, i.e. orthogonal then we have $|<\varphi_1|\varphi_2>|^2 = 0$, and the overlap is 1 when these states are perfectly indistinguishable, hence we can measure [20], first of all, the average time taken for one state (say in our case this is one state initially formed of QZE or collective actualization) to transform into another in general non-orthogonal state before decoherence takes over. This can be shown as the time average of $1 - |<\varphi_t|\varphi_{t'}>|^2$. Here t and t' means these states are emerging in different moments. Second, we can also measure in the same vein the amount of state evolution which has happened in any time interval, generally we have the measure $1 - \frac{1}{T^2}\iint |<\varphi_t|\varphi_{t'}>|^2 \, dt'dt$ where the limits of integrals are between 0 and T.

More on time measurements in our experiments: if we prepare systems in inter or intra entangled states, then we need to be able to define time 0 and the interval between that and the end of the decoherence process (on the order of femto-seconds or nano seconds), since this should be the time interval in which a specific basis emerges through the process of collective actualization. Then decoherence takes over, which might allow only the so-called pointer bases to survive. Here we can use the above defined formulas for measuring the amount of quantum evolution within that time scale. We can get some reference from the experiments on light harvesting molecules, where the time scale for decoherence is on the order of femtoseconds. Recent studies observe that QZE can be achieved by a series of arbitrary weak measurement [25], hence in the collective actualization process which is approximated by the many body QZE we may assume series of WMs happing between the bodies.

In the standard measurement scenario, we start with $|\Psi>$ as the pure state of the system before measurement, for convenience we assume the state can be expanded in computational basis. We can only calculate the probabilities of getting a definite value m, say prob (m), which is one Eigen value. In general we can start with positive operators, M, which are defined as $\sum_m MM^D = I$, where D is a symbol we use to signify the Hermitian conjugate operation. We will have Born's probability to get a specific Eigen value, $p(m) = <\Psi|MM^D|\Psi>$ for the mth eigenvalue. If the M s are also projective operators then $MM^D = M$, and we will get a projective measurement. Again the state of the system after measurement in general will be $M|\Psi>$. Call that $|\Psi m>$.

Say we start with $|\Psi> = 1/\sqrt{2}[|0> + |1>]$, where these states are the eigenstates for the Pauli matrix S in the Z direction $S(z) = \begin{pmatrix} 1 & 0 \\ 0 & -1 \end{pmatrix}$. A simple projection measurement on the state would be either via M = |0><0| projection on to |0> or M= |1><1| projection onto state |1>.

We can also measure the probability of the outcome of measurements by using POVM. With a slightly different completeness criterion: $\sum_m E_m = I$, where p(m)= <Ψ|$E_m$|Ψ>
We propose using weak measurements (WM), [26,27,28], to assess the emergence of a basis. This requires creating the initial state at T =0 in a chosen mixed basis, then assessing the possibly increasing amplitude for one of the bases over the subsequent decoherence process.

In this framework all systems are quantum, so generally, interactions are between quantum systems (in concordance with RQM). In the first step we set up a weak coupling between the target quantum system and the quantum measuring device. (Here we recall the collective actualising set of particles are peptides, so any one of the atoms-including isotope labelled atoms- in the set can be an observer.) In the second step we measure strongly with the quantum measurement device. The outcome of projective measurement on the quantum measuring device is the weak measurement (WM) outcome.

This weak measurement outcome assesses the amplitude for the bases used in the weak measurement. By this means, we can assess whether the amplitude for any basis increases or not over the interval from the moment , T = 0, of preparation in the chosen mixed basis, position – spin, position- polarization, or spin – polarization.

Using the above combination of weak coupling followed by strong measurement in a chosen basis following  T 0, it is possible to assess: i. The emergence of an amplitude for a given basis. Ii. The consequences for the emergence of an amplitude for one basis of a strong measurement of the other mixed basis.  Strong measurement of the second basis may have no effect on the emergence of the first basis, enhance or inhibit that emergence.

We note again that the emergence of a basis can be assessed in any one of the 10 isotope labelled atoms located arbitrarily in our test peptide.

 Mathematical necessary condition for a measurement to be weak is that standard deviation of the measurement has to be larger than the differences in the Eigen values of the system.

The capacity to carry out the experiments above depends on the ratio of decoherence and Quantum Zeno effect. By using quantum ordered, critical or chaotic peptides we can tune decoherence time from a  very slow process extending over a nano second for ordered and critical peptides to a short femtosecond decoherence time for chaotic peptides.

3.2.1 Experiments with Mixed Basis Entanglements

As mentioned earlier quantum variables can be prepared in mixed bases, for example entanglement between two variables between spatially separated quantum systems (polarization of one photon with spin of another, for example), or entanglement between two degrees of freedom of the same system (intra system entanglement). Such systems cannot be considered to have a definite 'identity', but measurements can be well performed on different pairs of entangled variables in two mixed bases. Also, there can be protocols built where there can be swaps between intra and inter system entanglements.

In our experimental set-ups we can prepare peptides/ amino acid molecules in such entangled states. Particularly, we assume a maximally entangled N systems such that each of the N variables is further represented in a mixed basis state, i.e., for each individual variable there is no definite basis. And for the entangled set of N such individual variables also there is no definite basis. Overall, we have both intra system entanglement and inter system entanglement.

We can set up an experiment such that each of N variables is prepared in the same two mixed bases entangled, say position and polarization, hence if a weak measurement is performed in one of the bases of an individual of N, then due to intersystem entanglement, same basis for the N-1 elements would emerge. Again, we may again deploy the measures of different ratios we have mentioned earlier.

But we can also prepare a system with some pairs of variables sharing two mixed position spin bases, while other pairs of variables share two mixed position polarization bases, while yet other pairs share two mixed spin polarization bases. Using at least 10 stable isotopes, see next, we can entangle a set of atoms in arbitrary combinations of the three different mixed bases and study their detectable behaviors.

The combinatorial possibilities for different patterns of entanglement of our isotope labeled atoms is very large. As noted, there are 3 stable isotopes of carbon, nitrogen, oxygen, and 5 for sulfur. One of each of the three is the normal isotope, so is in all the amino acids of our test protein. The other two isotopes of C, N, and O are not. So, we have two detectable isotopes each of C, N O, and 4 detectable isotopes of sulfur. Again, as noted, that is 10 detectable signals that we can put into any atoms in the peptide length N.

If we can arbitrarily entangle these 10 atoms, there are [10 choose 2] = 45 pairs of labeled atoms. For each pair of such atoms the following entanglements are possible: i. not entangled. ii entangled in any one of the three possibilities: position-spin, position – polarization, spin – polarization.  iii. The two isotope-labeled atoms can also be entangled with any two entangled bases: position-spin and position-polarization; position-spin and spin-polarization…There are (3 choose 2) = 6 such choices of two pairs of mixed bases. iv. Any pair of labeled atoms can be entangled with all three entangled bases. There is only one choice, (3 choose 3) of all three mixed bases.

In sum, any of the 45 pairs of labeled atoms can be in 11 different entanglement relations. Therefore, using only pairs atoms that are both labeled, there are $11^{45}$ different patterns of entanglement among 10 labeled atoms at different specific locations in the test peptide. This should allow detailed assessment of basis emergence as a function of temperature, N, and the ratio of QZE to decoherence.

One of the central strengths of the experiments we are suggesting is a plethora of different classes of entanglements which can result out of mutual actualization processes; it is well known in the literature [30] of multi-particle entanglement that when we move from two Qubits to three Qubits entanglement states, we have different classes of entanglements generated, for example W and GHZ. Along with these different classes we have both intra and inter entanglements.

**Conclusions**

It is widely supposed that as the number of atoms of a system increase it should become more 'classical'.  Well established work has studied aspects of classicality as the number of atoms increase. Buckyballs with C 60 still show interference.

We propose here these six new ideas: i. Collectively Actualizing Sets, whose variables interact with one another and thereby actualize one another. ii. Such collective interaction induces a symmetry breaking among two or more mixed bases to a single preferred basis. iii. An initial consideration of the joint effects of decoherence and internal Quantum Zeno Effects within such a CAS upon the emergence of the classical world. iv. The use of quantum ordered, critical, and chaotic peptides of lengths from one up to 100 amino acids or more as the test objects. v. The use of the three stable isotopes each of oxygen, carbon, nitrogen and five of sulfur to uniquely and identifiably isotope label any ten atoms placed in arbitrary positions within the test peptide. v. The use of mixed base entanglement among the pairs: position – spin, position – polarization, spin – polarization, momentum – spin, and momentum – polarization to study symmetry breaking between these two bases within a collectively actualizing set. vi. The use of weak measurements of isotope labeled atom in such peptides to assess the emergence and stability of one or the other among each of these five pairs of entangled bases.

We hope our proposal is seen as a continuation of a long tradition. The basic concepts seem reasonable. Creating a real mathematical framework is a large further task. So is assessing the feasibility of the proposed experiments.

A possible alternative mathematical framework: our framework of mutually collectively actualizing set is close to a very recently proposed framework for relational QM, Fact-Nets (4). The main proposal for fact-net framework is to recover the standard conditional probability measures in QM without having any quantum state as a physical entity, which is also the central proposition of relational QM, since the founders have thought that the major confusion in interpretations of QM has been due to placing any ontological weigh on the wave function. Fact-nets also do not need the Hilbert space formalism, but central features of such a formalism can be recovered. Here we are pragmatic about the interpretation problem, and mention that in subsequent progress with our project we might use Fact-nets as a possible coherent mathematical framework.

**Mathematical Appendices**

I. Here we discuss some further details of weak measures:

Describing the 'measuring device': we need to describe the wave function of the measuring device in a specific basis, say position basis, this is before the strong measurement on the measuring device, say $|\emptyset> = \int \emptyset(x)|x> dx$, where we define, we also define a position operator, $X|x> = x|x>$, $\emptyset(x)$ *such that its normally distributed around* $0$, *with a* $\sigma$. We later strongly measure $|\emptyset>$ to get a reading on the device which is the outcome of the WM.
We need a conjugate operator to X, say P, s.t. $[X,P] = ih/2\pi$
System/ body on which the measurement happens: we can decompose the state of the system in a given Eigen basis corresponding to a Hermitian operator say A acting on the system. Such that $A|a_k> = a_k|a_k>$, hence for the system's state $|\Psi> = \sum_k a_k |a_k>$.
Hence we consider the interaction Hamiltonian for the dynamics. This is between the system and the measuring device:
$H(int) = g(t) A \otimes P$. Where g(t) is the coupling impulse function $\int_0^T g\, dt = 1$
For the 'measurement' process then, the vector of relevance is $|\Psi> \otimes |\phi>$.
Then we have the dynamics of this weakly coupled state $e^{-iHt/h} |\Psi> \otimes |\phi>$.
Now we need to compare the variances of the wave functions, iff ϕ has a larger variance than the Ψ the waves would overlap and we have a scenario ready for weak measurement, otherwise we will have strong measurement.

The next step is a strong projective measurement on the 'measuring' device, which reveals the information about the initial system/ body's state with a slight bias.

Two state vector formalism

This is a formalism for describing WM and post selections. Introducing 'post selection' in the framework of WM results in various strange results, even negative probabilities. Say we prepare an ensemble of systems prepared in the state $|\varphi_{in}>$, then we weakly measure such an ensemble by a device such that the initial state of the device is $|\emptyset(x)>$, and the interaction Hamiltonian is $H_{int} = g(t)A \otimes P$ where A is a Herm operator on the S and P is the conjugate operator on device.

Now if we want to have the amplitude for the final state vector $|\varphi_{fin}>$, we can compute it by the transition probability rule. This means that perform measurement with H with a no of copies of the system, and choose only those results which have a state in the direction of fin.

Create operators $P_1 = |\varphi_{fin}><\varphi_{fin}| \otimes I_d$ and a sum form operator.
Then with the help of such operators do a strong measurement on composite state $e^{-iHt/h}|\varphi_{in}> \otimes |\emptyset(x)>$, which is $|\varphi_{fin}><\varphi_{fin}|e^{-iHt/h}|\varphi_{in}> \otimes |\emptyset(x)>$,
We also assume that $P_d$ has a lower variance as compared to $X_d$, hence weak coupling between system and device is possible.
Now the above vector approximates to $|\varphi_{fin}><\varphi_{fin}|(1 - iA \otimes PT/h)|\varphi_{in}> \otimes |\emptyset(x)>$
The above can be simplified as $|\varphi_{fin}> \otimes <\varphi_{fin}|\varphi_{in}> \left(1 - i<A>\frac{PT}{h}\right)|\emptyset(x)>$
Where $<> = <\varphi_{fin}|A|\varphi_{in}>/<\varphi_{fin}|\varphi_{in}>$
Now we can compute probability of post selection $<\varphi_w|P^D P|\varphi_w>$
, where P is $P_1$ actually .
Where $\varphi_w$ is defined early as the state: $e^{-iHT/h} |\varphi_{in}> \otimes |\emptyset(x)>$, hence if we plg in all these in the above expression we get P ()= $|<\varphi_{fin}|\varphi_{in}>|^2$

Weak measurements as universal POVM are general measurements: may capture many phenomena not revealed by projection measures: extra randomness in the measures or incomplete information in measures. We start with a density matrix description of initial state which undergoes random updating:
$\rho$ to $\rho_j$ Such that $\rho_j = \frac{P_j^+ \rho P_j}{tarce\ (P_j^+ \rho P_j)}$, where the denominator is the probability of the $j^{th}$ outcome.
Now a unitary transformation can be decomposed to sequence of weak unitary transformations.

Any generalised measurement as a sequence of weak measurements. Weak measures can be termed as those whose measures do not impact the initial state significantly.

There are other definitions for example measure which generates large change in state with a small probability.

Here we see $P_j = q_j(I + \epsilon_j)$, where q and epsilon are operators such that q (0,1) and epsilon is an operator with a very small norm. Weak measurements can be found in systems under continuous monitoring.

II. Here we describe the 'mixed basis entanglement':

There is a strong literature in quantum as well as classical optics (where we still consider Maxwell field equations rather than any quantum degrees of freedom, for example light quanta or photon) where states can be produced which expresses entanglement between multiple degrees of freedom. One such case is path-polarization entanglement, such states have been found to violate Bell or CHSH inequalities [33]. Hence such states cannot be considered to have any determinate basis.

Some authors [34] hold that classical entanglement is based on intra system or such path-polarization type entanglements, whereas genuine quantum entanglements are inter-system, for example EPR pairs.

III. Non-Hermitian Hamiltonian:

Earlier we mentioned Ising or a possible Spin Glass Hamiltonian for approximating the dynamics of our experimental peptide framework. Evidence supports the use of tunable rugged spin glass models, called NK models, for proteins [29]
Given a choice of Hamiltonian [31], we observe that overall there are two broad choices for describing an open system Markovian dynamics. One, where we can use GKSL master equations, under several assumptions, or where we use an effective non-Hermitian Hamiltonian. In the literature of many body QZE, non-Hermitian Hamiltonians have been used.

A substantial literature starting from Bender [32] onward has observed that Non-Hermitian Hamiltonians can also exhibit real Eigen values, provided PT (parity and time reversal) symmetry is embedded in such a formation. Exceptional points emerge in such Non-Hermitian dynamics where given PT symmetry Eigen values change from real to complex in general, a phase transition from so called unbroken PT symmetry to broken PT symmetry. The overall assumption of such dynamics is that though the underlying full Hamiltonian is Hermitian, the effective system Hamiltonian can be treated as non-Hermitian.

IV. Bohmian mechanics: basic quantum hydrodynamics approach

Modern renditions of Bohmian mechanics (for example here [36]) presents it as a first order theory, where velocity, which is the rate of change of positions, is fundamental. Velocity is provided by the so-called guiding equation. Second-order concepts like force or acceleration do not contribute in this version. However, in the original version, Bohm conceived of a second order theory, where forces derive from a non-local quantum potential. The main technique is to rewrite wave function in the polar form, which is $\varphi = R \exp 2\pi i S/h$ where R and S are real valued functions. Then re-writing the Schrödinger's equation in terms of these new variables, one obtains two coupled equations, one, a continuity equation $\rho = R^2$, and another, a modified Hamilton-Jacobi equation for S. the modified equation has an extra potential term $U = -\sum_k (h^2/2m_k)\partial_k^2 R/R$, which is termed as the quantum potential. Particle trajectories are then shown to be resulting from quantum potential, in addition to the usual forces.

**Acknowledgements**

We thank Professor Partha Ghose for helpful insights and discussions.

Authors did not receive any funding for the work.

The authors have no conflicting issues and have the right to publish this material.

**References**

1. Penrose, R., Shimony, A., Cartwright, N. and Hawking, S., 2000. The large, the small and the human mind. Cambridge University Press.
2. Xavier, J.[1*], Hordijk, W.[2], Kauffman, S.[3], Steel, M.[4], Martin, W.F.[1] (2020). Autocatalytic chemical networks at the origin of metabolism. March 11, 2020, Proc Roy Soc B.
3. Rovelli, C., 1996, Int. J. Theoretical Phys. 35, 1637
4. Laudisa, F. and Rovelli, C. "Relational Quantum Mechanics," The Stanford Encyclopedia of Philosophy (2019), Edward N. Zalta, ed.
5. Pienaar, J., 2021. QBism and Relational Quantum Mechanics compared. Foundations of Physics, 51(5), pp.1-18.
6. Martin-Dussaud, P., Carette, T., Głowacki, J., Zatloukal, V. and Zalamea, F., 2022. Fact-nets: towards a mathematical framework for relational quantum mechanics. arXiv preprint arXiv:2204.00335.

7. Facchi, P. and Pascazio, S. 2008 J. Phys. A: Math. Theor. 41 493001
8. Nourmandipour, A., Tavassoly, M.K. and Bolorizadeh, M.A., 2016. Quantum Zeno and anti-Zeno effects on the entanglement dynamics of qubits dissipating into a common and non-Markovian environment. JOSA B, 33(8), pp.1723-1730.
9. Zurek, W. H., Phys. Rev. D 26 (1982) pp.1862–1880
10. Page, D.N., 2021. Does Decoherence Make Observations Classical? arXiv preprint arXiv:2108.13428.
11. Yan, B. and Zurek, W.H., 2021. Decoherence factor as a convolution: an interplay between a Gaussian and an exponential coherence loss. arXiv preprint arXiv:2110.09463.
12. Kauffman, S.A. and Weinberger, E.D., 1989. The NK model of rugged fitness landscapes and its application to maturation of the immune response. Journal of theoretical biology, 141(2), pp.211-245.
13. Vattay, G. Kauffman, S., Niiranen, S. (2012). Quantum biology on the edge of quantum chaos. Physics arXhiv: http://arxiv.org/abs/1202.6433
14. Vattay, G., Salahub, D., Csabai, I., Nassimi, A, and Kauffman, S. (2015). Quantum criticality at the origin of life, 7th International Workshop DICE2014 Spacetime – Matter – Quantum Mechanics IOP Publishing Journal of Physics: Conference Series 626 (2015) 012023 doi:10.1088/1742-6596/626/1/012023
15. Po-Wen Chen, Dong-Bang Tsai, and Bennett, P. Phys. Rev. B 81, 115307.
16. Zurek, W.H., 2021. Emergence of the Classical from within the Quantum Universe. arXiv preprint arXiv:2107.03378
17. Xie, H., Jiang, F., Tian, H., Zheng, X., Kwok, Y., Chen, S., Yam, C., Yan, Y. and Chen, G., 2012. Time-dependent quantum transport: An efficient method based on Liouville-von-Neumann equation for single-electron density matrix. The Journal of Chemical Physics, 137(4), p.044113.
18. Paz-Silva, G.A., Rezakhani, A. T., Dominy, J. A. and Lidar, D. A. Phys. Rev. Lett. 108, 080501.

19. Zurek, W. H. 2018 Quantum theory of the classical: quantum jumps, Born's Rule and objective classical reality via quantum Darwinism Phil. Trans. R. Soc. A. 3762018010720180107.

20. Biella, A. and Schiró, M. "Many-body Quantum Zeno effect and measurement-induced subradiance transition." Quantum 5 (2021): 528.

21. Amit, D. J., Gutfreund, H. and Sompolinsky, H. "Spin-glass models of neural networks." Physical Review A 32.2 (1985): 1007.

22. Chiarelli, P. and Chiarelli, S., 2021. Stability of Quantum Eigenstates and Collapse of Superposition of States in a Fluctuating Vacuum: The Madelung Hydrodynamic Approach. European Journal of Applied Physics, 3(5), pp.11-28.

23. Restrepo-Pérez, L., Chirlmin Joo, and Dekker, C. "Paving the way to single-molecule protein sequencing." Nature Nanotechnology 13.9 (2018) pp. 786-796.

24. Gatto, E., Toniolo, C. and Venanzi, M. "Peptide Self-Assembled Nanostructures: From Models to Therapeutic Peptides." Nanomaterials 12.3 (2022) p. 466.

25. Scholes, G.D., Fleming, G.R., Olaya-Castro, A. and Van Grondelle, R., 2011. Lessons from nature about solar light harvesting. Nature chemistry, 3(10), pp.763-774.

26. Dominy, Jason M., et al. "Analysis of the Quantum Zeno effect for quantum control and computation." Journal of Physics A: Mathematical and Theoretical 46.7 (2013): 075306.

27. Sgroi, P., Palma, G. M. and Paternostro, M. "Reinforcement learning approach to nonequilibrium quantum thermodynamics." Physical Review Letters 126.2 (2021): 020601.

28. Aharonov, Y. and Vaidman, L., 2008. The two-state vector formalism: an updated review. Time in quantum mechanics, pp.399-447.

29. Vaidman, L., 2018. Comment on 'Two-state vector formalism and quantum interference'. Journal of Physics A: Mathematical and Theoretical, 51(6), p.068002.

30. Kauffman, S. A. and Weinberger, E. D. (1991). The NK model of rugged fitness landscapes and its application to the maturation of the immune response.

Molecular Evolution on Rugged Landscapes: Proteins, RNA and the Immune System, vol. IX (ed. Perelson, A. S. and Kauffman, S. A.), pp. 135-177. Reading, MA: Addison Wesley.

31. Patra, S., and Ghose, P. "Measurement, Lüders and von Neumann projections and non-locality." Pramana 96.1 (2022) pp. 1-10.
32. Bagarello, F., Passante, R. and Trapani, C. "Non-Hermitian Hamiltonians in quantum physics." Springer Proceedings in Physics 184 (2016).
33. Bender, C. M. "Making sense of non-Hermitian Hamiltonians." Reports on Progress in Physics 70.6 (2007) p. 947.
34. Ghose, P. and Mukherjee, A. (2014). Entanglement in classical optics. Reviews in Theoretical Science, 2(4), pp.274-288.
35. Khrennikov, A., (2020). Quantum versus classical entanglement: eliminating the issue of quantum nonlocality. Foundations of Physics, 50(12), pp.1762-1780.
36. Bohmian Mechanics (Stanford Encyclopedia of Philosophy) https://plato.stanford.edu/entries/qm-bohm/